\documentclass[twocolumn]{aastex63}
\usepackage{subfigure}
\usepackage{color}
\usepackage{xcolor}
\usepackage{amsmath}
\usepackage{mathrsfs}
\usepackage{multirow}
\usepackage{soul}
\usepackage{amssymb}
\usepackage{hyperref}
\usepackage{bm}

\def\degree{{}^{\circ}}

\received{...}
\revised{...}
\accepted{...}

\def\be{\begin{equation}}
\def\ee{\end{equation}}
\def\bd{\begin{displaymath}}
\def\ed{\end{displaymath}}
\def\ba{\begin{aligned}}
\def\ea{\end{aligned}}
\def\bh{M_{\bullet}}

\def\msun{M_{\odot}}

\begin{document}

%\title{Probing the Spinning of massive black hole by X-ray Quasi-periodic Eruptions due to 
%Star Disk collision and the application to GSN069}

\title{X-ray Quasi-periodic Eruptions driven by Star-Disc Collisions : 
Application to GSN069 and Probing the Spin of Massive Black Holes}
\author{Jingtao Xian}
\affiliation{School of Physics and Materials Science, 
Guangzhou University, 
510006 Guangzhou, China}
\author{Fupeng Zhang}
\correspondingauthor{Fupeng Zhang}
\affiliation{School of Physics and Materials Science, 
Guangzhou University, 
510006 Guangzhou, China}
\affiliation{Key Laboratory for Astronomical Observation and Technology of Guangzhou, 510006 Guangzhou, China}
\affiliation{Astronomy Science and Technology Research Laboratory of Department of Education of Guangdong Province, Guangzhou 510006, China}
\email{zhangfupeng@gzhu.edu.cn}
\author{Liming Dou}
\affiliation{School of Physics and Materials Science, 
Guangzhou University, 
510006 Guangzhou, China}
\affiliation{Key Laboratory for Astronomical Observation and Technology of Guangzhou, 510006 Guangzhou, China}
\affiliation{Astronomy Science and Technology Research Laboratory of Department of Education of Guangdong Province, Guangzhou 510006, China}
\author{Jiasheng He }
\affiliation{School of Physics and Materials Science, 
Guangzhou University, 
510006 Guangzhou, China}
\author{Xinwen Shu}
\affiliation{Department of Physics, Anhui Normal University, Wuhu, Anhui 241002, China.}
\begin{abstract}
X-ray quasi-periodic eruptions (QPEs) are discovered recently in active galaxies
with unknown driven mechanism. 
 Under the assumption that QPEs are caused by star-disc collisions, we 
adopt full relativistic method and show that both the orbital parameters of the star and also the mass and spinning 
of the MBH can be revealed by using the time of arrival (TOA) of the QPEs. 
By applying the model to the observed QPEs of GSN069, we find that the star is in a near-circular 
orbit ( {$e_\bullet=0.05^{+0.02}_{-0.02}$}) with semimajor axis of 
 {$\sim 365^{+54}_{-49}r_{\rm g}$} around a MBH with  {$\bh=3.0^{+0.9}_{-0.6}
\times10^5\msun$}. The alternative short and long recurring time
of the QPEs of GSN069 can be well explained by the small eccentricity and the orbital precession of the
star. { We find that the QPEs of GSN069 are possibly driven by a 
striped stellar core colliding with accretion disc after partial tidal disruption 
event around the massive black hole (MBH).}
For GSN069-like galaxies, if continuous X-ray monitoring of QPE events can be accumulated with
 uncertainties of TOA ${\lesssim 100-150}$s, the spin of massive black hole can 
be constrained by fitting to QPEs. Our results show that the timing of QPEs can provide 
a unique probe for measuring the spinning of MBH and tests of no-hair theorem. 
\end{abstract}


\section{Introduction}
X-ray quasi-periodic eruptions (QPE) are rapid and extremely intensive bursts of X-ray emission
which repeat about every few hours from regions around massive black holes (MBH) in galactic nuclei.
QPEs have been found recently from the active galactic nucleus of GSN069~\citep{2019GSN069QPE} 
and galaxy J1301.9+2747 \citep{2020A&A...636L...2G}. More recently, QPEs are detected 
in the nucleus of two previously quiescent galaxies~\citep{2021Natur.592..704A}. Multiple 
epoch of observations suggest that QPEs may last from at least months to 
decades~\citep{2019GSN069QPE,2020A&A...636L...2G}. Currently it is still unclear of 
the underlying physical mechanism
that drive the burst of these events. Some of the possibilities are the limit-cycle oscillations induced
by instabilities of the accretion flow~\citep{1974ApJ...187L...1L}
or an edge-on binary black hole gravitationally lensing the light from each others' accretion disc 
(AD)~\citep{2021MNRAS.503.1703I}. However, they are both found inconsistent with the properties of 
the eruptions~\citep{2021Natur.592..704A}. { Other possiblities are the falling 
clumps of partial TDE~\citep{2020ApJS..247...51C}, or the outburst arised due to the 
misalignment between the disc and the spin of the MBH~\citep{2021ApJ...909...82R}.}
One of the tempting mechanisms is that the X-ray QPEs are 
driven by an orbitor around the MBH that is with masses much smaller than the 
MBH~\citep{2021Natur.592..704A}. One of such examples is a white dwarf partially 
disrupted by MBH when crossing the periapse per orbit~\citep{2020MNRAS.493L.120K}.

It has been suggested that quasi-periodic flares may be produced by collisions between a star and 
an accretion disc surrounding MBH~\citep{1994ApJ...422..208K, 2004A&A...413..173N, 2010MNRAS.402.1614D}.
Due to collision on the disc, gas are shocked and subsequently a bright hot spot may appear and evolve on 
the disc~\citep{1983Ap&SS..95...11Z,1998I,2004A&A...413..173N,2016BHDI, 1999PASJ...51..571S,2019Sci...363..531P}.
The star-disc collision model may be able to explain the X-ray periodicity of galaxy RE 
J1034$+$396~\citep{2010MNRAS.402.1614D,2008Natur.455..369G}, and the 
optical variability of source OJ287~\citep{1996ApJ...460..207L,2010MNRAS.402.1614D}.
By adopting post-Newtonian methods, \citet{2008Natur.452..851V,2011ApJ...742...22V} show 
that the timing of these flares may be a useful probe for general relativity in strong field.

Here we study the timing of X-ray QPEs under the assumption that they are driven by the star-disc 
collisions. We use a full general relativistic numerical method~\citep[developed in][]{2015ZFP,2017ApJ...849...33Z} 
to simulate the orbit of a star and also the propagation of lights emitted from the disc to the observer. 
We then develop a numerical method which can extract the orbital parameters of the star 
from the timing of X-ray QPEs. We apply the method to GSN069 and set constraints the orbit of the star. { We also discuss the possiblity that the orbitor in GSN069 is a result of partial tidal disruption event. We further show that using QPE alone can set strong constraints on the spinning of the massive black hole if QPEs can be observed with high enough timing accuracy}. Details of the modeling and the results are shown 
in the following sections.

\section{Model and Methodology}
\label{sec:model_method}

\begin{figure*}
\centering
    \includegraphics[scale=0.37]{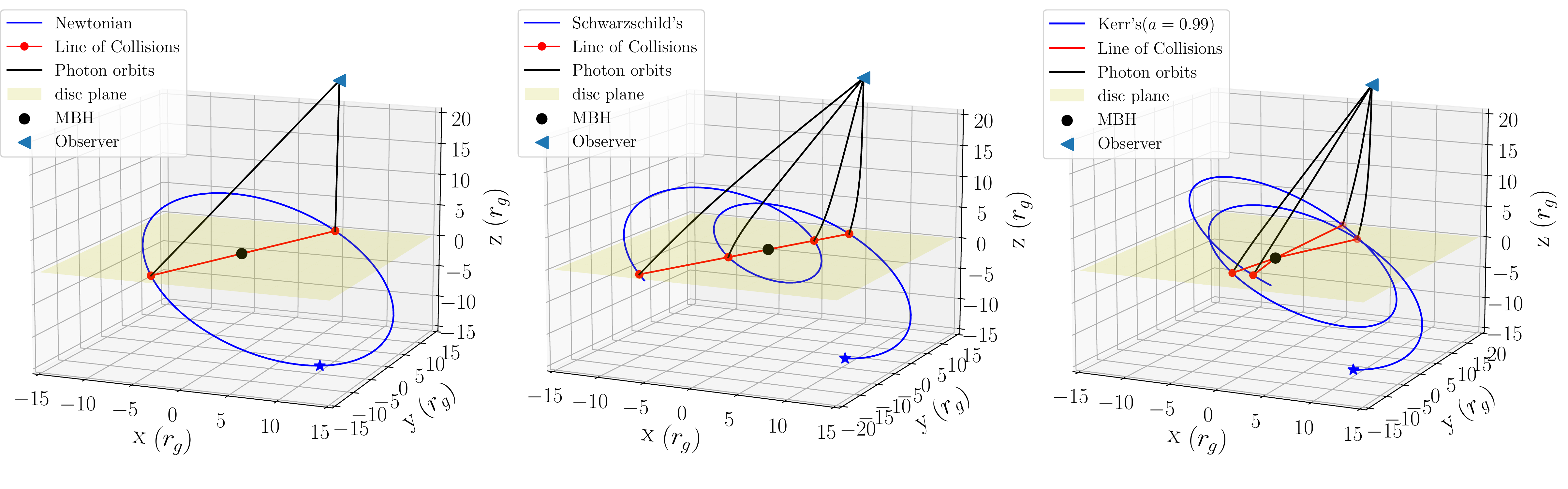}
    \caption{ Illustration on the trajectories of the orbitor and the photons 
    in Newtonian case (left panel) and relativistic cases for a Schwarzschild MBH (middle panel) and 
    a spinning MBH (right panel). For illustration purpose, the observer is located at distance of 
    $r=30r_g$ from the MBH (for simulations performed in this work, it is actually at $r=10^7r_g$). Note that the ascending node or descending node of the orbitor 
    (the red solid line) remain constant for a Schwarzschild MBH, but change in each revolution for a spinning MBH due to the precession of the orbital plane of the orbitor. 
    }
    \label{fig:ill}
\end{figure*}

The prime assumption is that the X-ray eruptions are due to super-sonic collisions between a star\footnote{ We will show later that the orbitor is actually more consistent with a remnant core of 
a red giant after partial tidal disruption event. Nevertheless, we generally call the orbitor  ``a star'' 
across the paper.} bound 
to the MBH and a standard geometrically thin AD, similar to those in~\citet{1995MNRAS.275..628R} and \citet{ 
2010MNRAS.402.1614D}. 
Here we adopt full general relativistic numerical framework developed in~\citet{2015ZFP} 
(See also \citet{2017ApJ...849...33Z}) to calculate the orbit of 
the star and the propagation of lights from the disc to observer under Boyer-Lindquist coordinates. 

Suppose that the peak of each QPE corresponds to the time when the star
intersects with the midplane of the disc. {
At each time of intersection, we use ray-tracing technique to trace back a number of 
photons from a distant observer until we find the one that hits on the position within distance of 
$<10^{-8}r_\bullet$ from the star, where $r_\bullet$ is the distance of the star to the MBH.
The ray-tracing is fast and accurate as we use Jacobian elliptic functions and Gauss-Kronrod integration 
scheme for the integration of motion equation of photons (For more details see~\citet{2015ZFP}). 
}

Then the time of arrival (TOA) $t_{\rm TOA}$ of 
the eruption due to collision measured in the observer's frame is then given by
\begin{eqnarray}
    t_{\rm TOA} = t_\star + t_{\rm prop}
\end{eqnarray}
where $t_\star$ is the coordinate time of the star when it intersects with the disc mid-plane 
and $t_{\rm prop}$ is the time of propagation { of the photon from the intersection to the observer 
found by the ray-tracing method.
}

{ For simplifity, }we ignore the secondary image of the eruption produced by  photons running 
around the other side of the MBH and twisting back to the observer, as usually the amplitude of 
these high order images are small~\citep{1994ApJ...422..208K,2010MNRAS.402.1614D}.
{ We also ignore the gravitational wave orbital decay and the possible deviations of the 
star orbit due to collision for now, as we find later that they 
can be safely ignored for GSN069 (See more details in Section~\ref{sec:possible_pTDE}).
}

{ Figure~\ref{fig:ill} illustrate examples of trajectories of orbitors and photons
in Newtonian case (left panel) and relativistic cases under Schwarzschild and spinning 
MBHs (middle and right panel). We can see that}
the orbit of the star intersects with the accretion disc twice per orbit, with the time of intersection 
separating with alternative short and long intervals if there is a non-zero orbital eccentricity.
In the meanwhile, the orbit of the star is precessing due to 
relativistic effects (Schwarzschild orbital precession and etc), 
thus the location of the intersection will change in each revolution. 
Due to the curved spacetime under Schwarzschild or Kerr metric, 
the propagating time of photons from the position 
of the intersection to the observer is also different. Combining these effects, 
the TOA of each eruption may then appear with irregular periods.

As the TOA of X-ray QPEs encode various orbital and relativistic effects, in principle the 
parameters of the MBH or orbit of the star should be reconstructed by the timing 
of QPEs. We then apply our method to the QPEs observed in GSN069, with more details 
shown in the following section. As we find that the current observations of GSN069 
are not able to recover the spin parameters of the MBH, we discuss the constraints 
of spin parameters from timing of QPEs for future observations in Section~\ref{sec:spin}.

\section{Extracting the orbit of star by X-ray timing of QPEs in GSN069}
\label{sec:res_GSN069}

\begin{figure*}
\centering
    \includegraphics[scale=0.45]{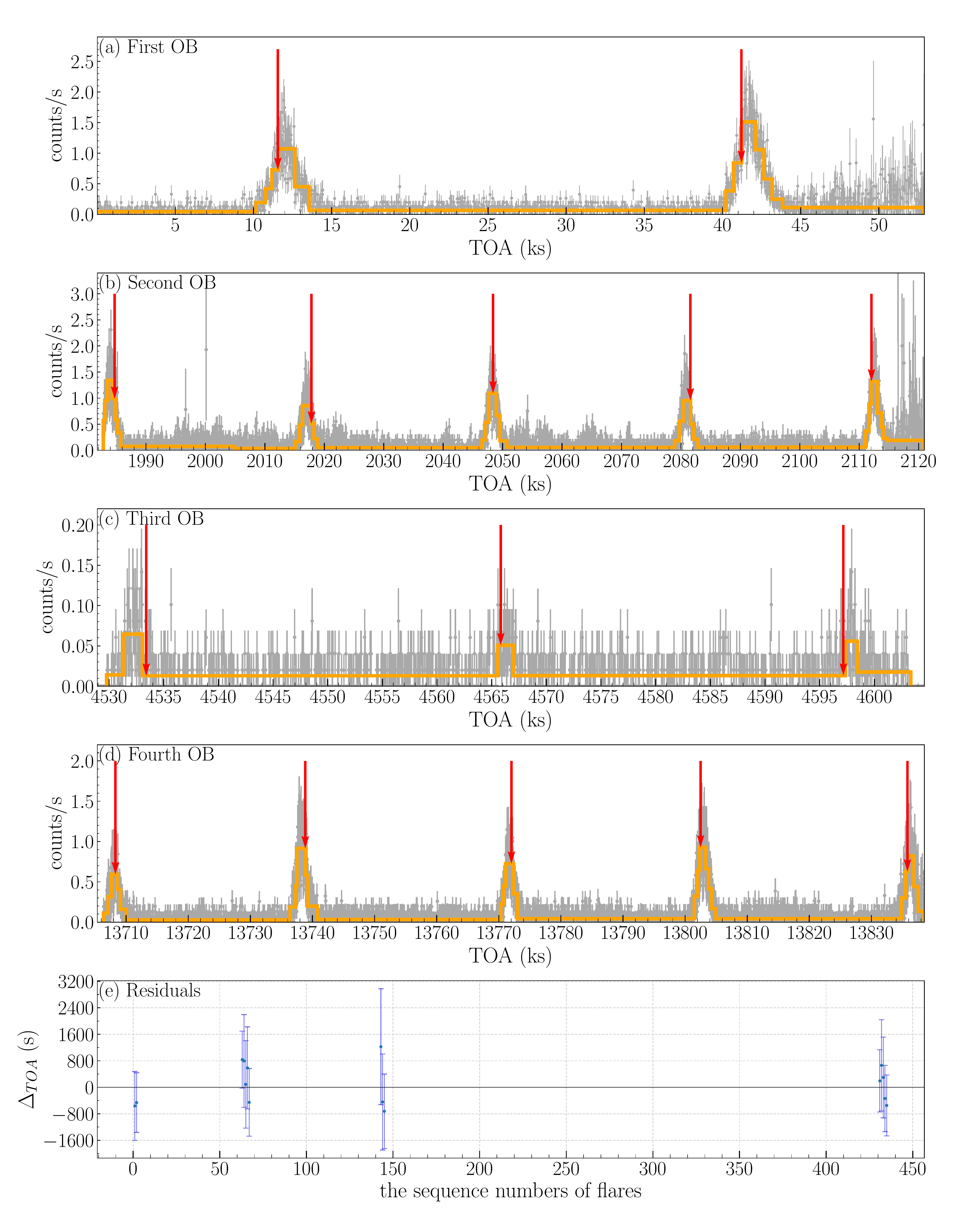}
    \caption{Panels (a)-(d): The observed X-ray flux of QPEs in GSN069~\citep{2019GSN069QPE} from December 2018 to 
    May 2019
    (grey dots with error bar) and the TOA of QPEs from the best-fit models (marked by the red arrows). 
    The orange solid lines show the optimal Bayesian block representations near each eruption.
    Panel (e) show the residual of the TOA between the model and the observation, where the error bars show
    the $2\sigma$ uncertainty of the TOA. The horizontal axis 
    of Panel (e) show the number of sequences of each eruption from the modeling.
    }
    \label{fig:gsn069mcmcres}
\end{figure*}

\begin{figure*}
\centering
    \includegraphics[scale=0.4]{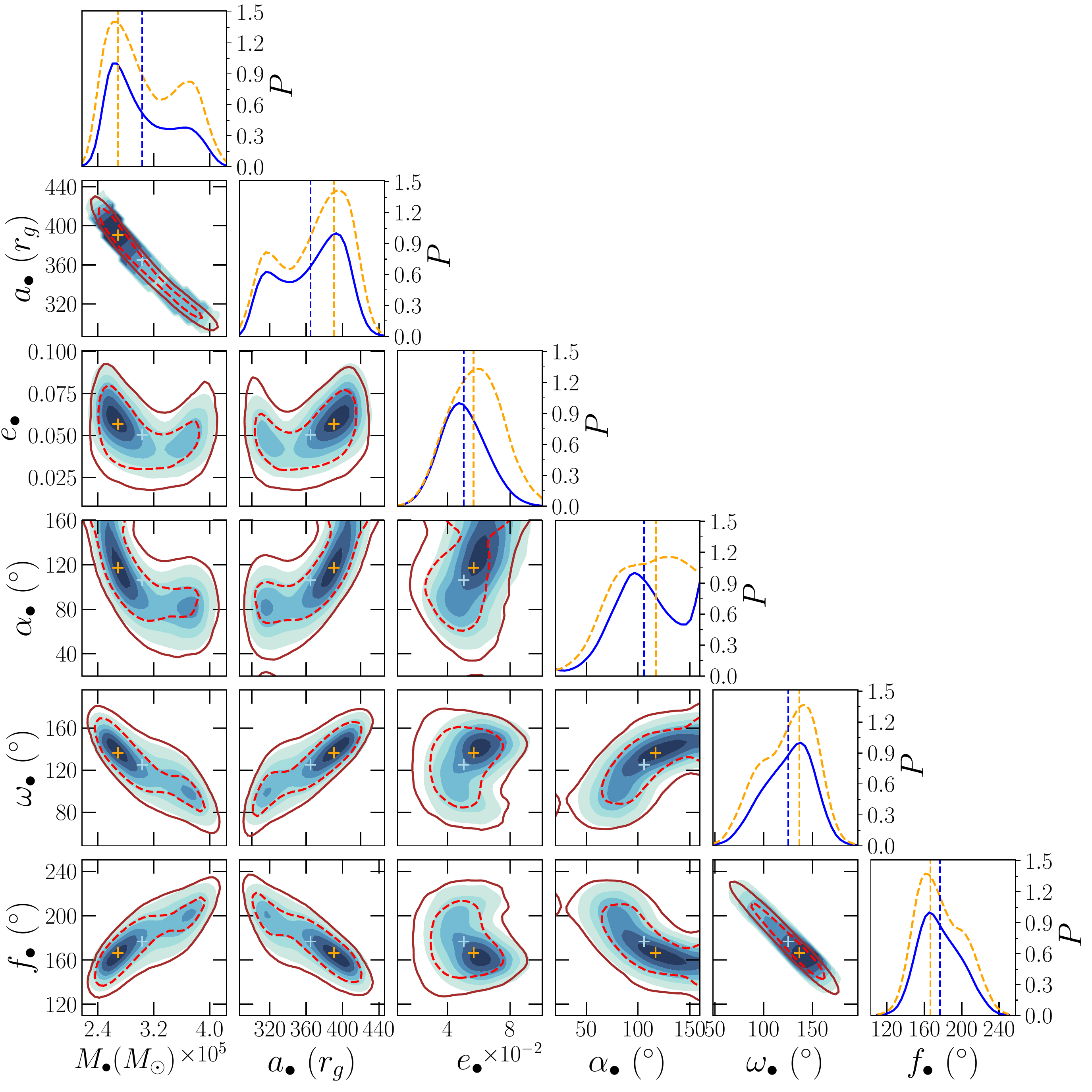}
    \caption{
      The reconstructed initial orbital elements of the orbiter
      by using X-ray timing of GSN069. 
      The color contour maps represent the mean likelihood of the MCMC sample, and the line contours represent
      the marginalized distribution, with the $1\sigma$ (dashed line) and $2\sigma$ (solid line) confidence level.
      The panels in triangle show the one-dimensional probability distribution. The mean-likelihood value and 
      those with minium $\chi^2$ of each parameters are marked by the light-blue and yellow cross in each panel, respectively.
     }
     \label{fig:gsn069mcmccontour}
\end{figure*}

%\begin{table}%[h]
%\centering
%\caption{Time of arrival of QPEs}
%\begin{tabular}{|c|c||c|c|}\hline
% TOA (ks) & 1$\sigma$ (s)  &TOA (ks) & 1$\sigma$ (s) \\ \hline
%      $ 12.12 $  & $ 520 $ &   $ 4566.27 $ & $ 725 $  \\ \hline
%    $ 41.67 $    & $ 450 $ & $ 4597.88 $&  $ 562.5 $ \\  \hline
%    $ 1983.89 $  & $ 430 $ & $ 13708.08 $&   $ 470 $  \\ \hline
%    $ 2017.04 $  & $ 700 $ & $ 13738.18 $&  $ 690 $   \\ \hline
%    $ 2048.3 $   & $ 660 $ & $ 13771.74 $&   $ 610 $ \\ \hline
%    $ 2081 $     & $ 620 $ & $ 13802.83 $ & $ 500 $  \\   \hline
%    $ 2112.51  $ & $ 510 $ & $ 13836.35 $ & $ 460 $ \\ \hline
%    $ 4532.17  $ & $ 875 $ &       &   \\ \hline
%  \end{tabular} 
%  \tablecomments{ the TOA of X-ray QPEs in GSN069  and the $1\sigma$ measurement errors
%  of each flare obtained by Bayesian blocks method~\citep{2013ApJ...764..167S}.
%  }
%  \label{tab:measure}
%\end{table}

\begin{table}[ht!]
\centering
\caption{Orbital elements reconstructed from QPEs in GSN069}
  \begin{tabular}{lccc}\hline
     paramaters            & range$^a$          & constraints$^b$                   &     best fit$^c$         \\ \hline \hline
     $ P_k [ks]$           & (0 $10^4$)         & $64.222^{+0.105}_{-0.073}$        &     $64.180$              \\ 
      $ \bh [M_\odot]$     & (0 $3\times10^8$)  &$3.03^{+0.92}_{-0.61}\times10^{5}$ &    $2.69\times10^{5}$     \\ 
      $ a_\bullet [r_g]$   & (100 700)          &  $364.8^{+54.0}_{-49.0}$          &   $390.4$               \\ 
      $ e_\bullet $        & (0 1)               &$0.050^{+0.030}_{-0.026}$         &   $0.057$                 \\ 
$\alpha_\bullet [\degree]$ & {( 20 160)$^{d}$}            &$106.0^{+54.0}_{-49.0}$            &   $117.1$                 \\ 
$ \omega_\bullet [\degree]$& (0 360)            &$125.1^{+42.5}_{-50.6}$            &   $136.4$                 \\ 
   $ f_\bullet [\degree] $ & (0 360)            &$ 176.7^{+45.9}_{-38.8}$           &   $166.5$                 \\ \hline
  \end{tabular} 
  \tablecomments{$^a$The prior ranges for parameters.\\
  $^b$The constraints of parameters from mean-likelihood in $2\sigma$ confidence level. \\
  $^c$The parameters with minium $\chi^2$ value ($\chi^2=16.2$). \\
  { $^d$Here $\alpha_\bullet$ should avoid angle near $0$ or $180$ degree 
  where the disc is edge on, which can cause difficulties in the light tracing 
  numerical methods.}
  }
  \label{tab:mcmcpars}  
\end{table}

\begin{figure*}[ht!]
    \centering
    \includegraphics[scale=0.28]{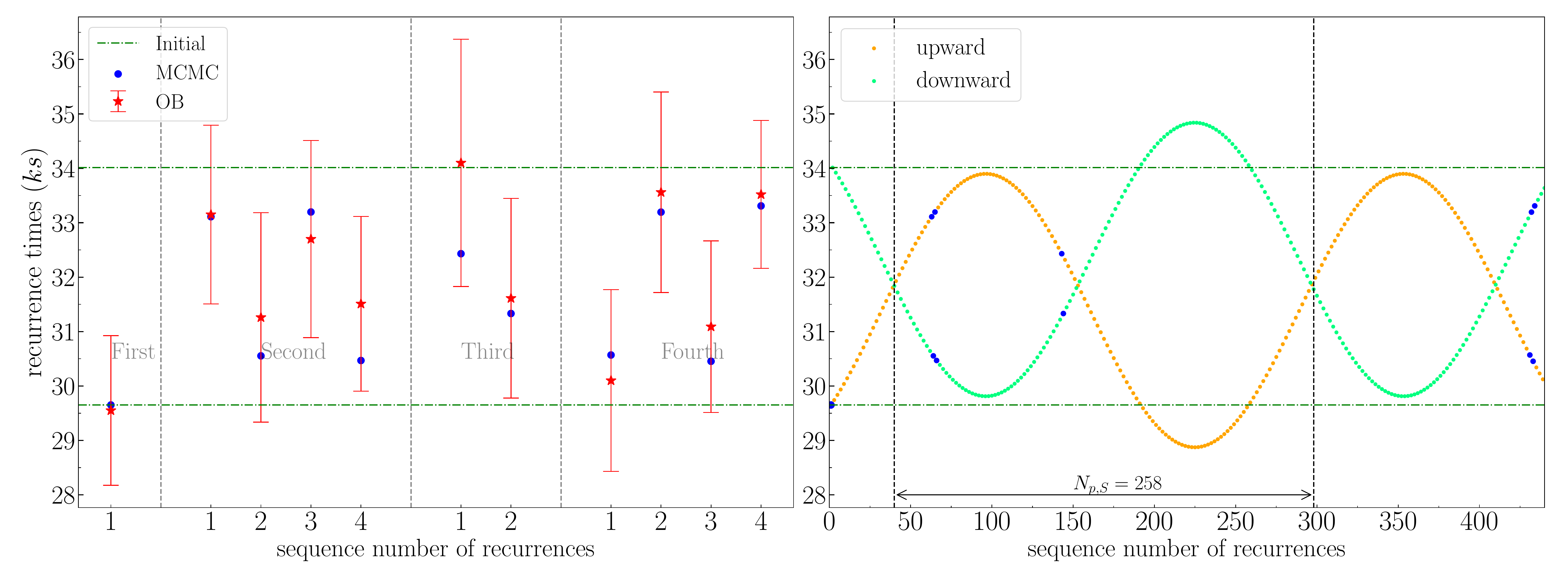}
    \caption{\label{fig:gsn069rec}
        The time interval between flares in each period of observation. { Left panel:} Data marked by red star 
        and the blue dot represents the observation of GSN069 and the corresponding best-fit results 
        in the MCMC modeling, respectively. The error bars show the $2\sigma$ error of the 
        uncertainties in TOA. { The dash-dot dark green lines ($34.01$ks and $29.65$ks) show the expected time interval in Newtonian case. Right panel: Time interval between QPEs for the best fit model for in total of 450 flares. Green (yellow) dots show the 
        interval between the moment that the orbitor cross the disc upward (downward) and 
        that downward (upward). }}
\end{figure*}
Multiple QPEs on Seyfert galaxy GSN 069 have been reported by~\citep{2019GSN069QPE}, 
which recurrent every $\sim 9$hr with duration of about $\sim 1$hr
around a MBH of mass $4\times10^5\msun$ according to the timescale~\citep{2019GSN069QPE}. 
The four epoch of observations collect up to $15$ QPEs in time span of about $\sim 160$ days. 

In principle, the MBH-star system contains in total of $9$ independent parameters:
$3$ independent parameters for MBH: mass ($\bh$),
the dimensionless magnitude of spin ($a$), the line of sight inclination of spin 
($i$, or the normal direction of accretion disc if $a=0$); 
$6$ parameters of the orbit of star that defined on the disc plane: 
the semimajor axis (SMA, $a_\bullet$), eccentricity ($e_\bullet$), inclination ($I_\bullet$), 
ascending node ($\Omega_\bullet$, defined respect to the projection line of sight on 
the disc plane.), argument of periapsis $\omega_\bullet$, and 
the true anomaly $f_\bullet$. Parameter $\bh$ can also be replaced by the Keplerian orbital period 
$P_k=2\pi(a_\bullet^3/\bh)^{1/2}$ as $\bh$ can be determined given $P_k$ and $a_\bullet$.

Due to the limited timing accuracy of each TOA, 
we found that the spin of MBH can not be reconstructed for GSN069, 
thus we fix $a=0$. If alternatively set $a=0.99$, we find that the 
constraints on other parameters are only weakly affected. 
The independent parameters of the model in case of a Schwarzschild black hole 
reduce to $6$~\citep{2010MNRAS.402.1614D}: 
$P_k$ or $\bh$, $a_\bullet$, $e_\bullet$, $\alpha_\bullet$, $\omega_\bullet$ and 
 $f_\bullet$, where $\alpha_\bullet$ is the angle between the vector 
pointing from the MBH to the 
ascending node of the star and the vector pointing from the MBH to the observer, 
and $\cos \alpha_\bullet=\sin{i} \cos{\Omega_\bullet}$.

We then use the Markcov-Chain Monte-Carlo (MCMC) method to reconstruct these parameters. 
As the observations are on discrete time periods with 
different number of eruptions, it is difficult to correspond an eruption in observation 
to that in simulation. Thus, here we evaluate the $\chi^2$ value by a method similar to those in the pulsar timing. 
Suppose that there are a sequence of QPEs indexed by $k=1,2,\cdots$ and the TOA of the $k$-th one is 
given by $t_{{\rm TOA},k}=\mathcal{F}(k)$. In observation, not all of them are covered. 
If there are $N$ discrete observations, and 
the number of QPEs in each one is given by $M_i$, $i=1,\cdots, N$, then $\chi^2$ 
is given by
\be\ba
    \chi^2_{d} &= \sum_{i=1}^N \sum_{j=1}^{M_i} \frac{(\mathcal{F}^{-1}(t_{{\rm TOA}, ij}) - (I_{i,j=1} +j - 1) )^2}{\sigma_{pij}^2},
\ea \ee
where
\be\ba
    \sigma_{pij} &= \sigma_{tij} \frac{\partial (\mathcal{F}^{-1})}{\partial t}\bigg|_{t=t_{{\rm TOA},ij}} \nonumber,
\ea\ee
and $\mathcal{F}^{-1}(t_{{\rm TOA},ij})$ is the inverse function which connects the TOA of the $j$-th flare
of the $i$-th observation to the index $k$.
$I_{j=1}$ is an integer which is closest to the index given by 
$\mathcal{F}^{-1}(t_{{\rm TOA},i1})$. $\sigma_{tij}$ is the measurement error of the TOA of the $j$-th flare of the $i$-th observation.

Due to the random brightness fluctuations of the QPE events, it is difficult to 
get directly the peak of each QPE from the observed lightcurves. 
We adopt the Bayesian block method~\citep{2013ApJ...764..167S,2010MNRAS.402..307V} 
which can split the time series to optimal segments
to help to identify the peak location of each QPE event. 
The analysis results for the observed QPEs of GSN069 are shown in Figure~\ref{fig:gsn069mcmcres}.
The TOA of each QPE is reasonably expected to be located at the center of the 
Bayesian blocks with maximum flux of a local time series, and 
the $1\sigma$ measurement error of TOA is assumed the half of the time span of that block.
The uncertainty of the TOA for QPEs in GSN069 are found between $430-875$s.
%The details of the TOA of each QPEs in GSN069 and the corresponding 
%measurement errors can be found in Table~\ref{tab:measure}.

%Here we assume that the peak of QPE is the TOA of the eruption when the star intersect with the 
%midplane of the disc. However, it is possible that there are irragular shifts of TOA due to the 
%complexities of the radiative processes. Such complexities may introduce an additional 
%systematic errors of the TOAs that can not be simply estimated in this work. 
%Thus, our above esimations provide only a rough estimation of both the peak 
%position and a lower limit of the uncertainties of TOA. 

The reconstructed parameters are shown in Figure \ref{fig:gsn069mcmccontour} and Table \ref{tab:mcmcpars}.
We find that the TOA of QPEs of GSN069 are consistent with an orbiter with SMA of  {$\sim 365r_g$}
and a small eccentricity of $ {\sim 0.05}$   
(according to the mean likelihood, marked by light-blue cross in each panel of Figure~\ref{fig:gsn069mcmccontour}). 
 The angle ${\alpha_\bullet\sim 106\degree}$ suggest that the light 
of sight direction is tilted to nearly perpendicular to the vector pointing from the MBH to 
the ascending node of the star intersect with the accretion disc. 

The best-fit value deviating slightly from those from mean likelihood in Table~\ref{tab:mcmcpars},
is due to two local minimum of $\chi^2$, as more clearly 
shown in Figure \ref{fig:gsn069mcmccontour}. The two local minimum is more apparent 
for {$\bh$} and $a_\bullet$, where one of them 
are the best-fit value at {$\bh=2.69\times10^5\msun$} and {$a_\bullet=390r_g$} (marked by yellow cross), and the other
are at {$\bh\sim3.8\times10^5\msun$} and $a_\bullet\sim 320r_g$. As the size of the MCMC chains is 
sufficiently large ($\sim 10^5$ accepted MC samples), the two local minimum is more likely due to the
fluctuations caused by small sample size of the QPE or some unknown systematics in the data.

Comparison of the observed QPEs of GSN069 and the TOA of each 
eruption in the best-fit model are shown in Figure \ref{fig:gsn069mcmcres} (marked by red arrows). 
The TOA of each QPE from observations 
are well consistent with those from simulations within $2\sigma$ errors.

The time intervals between QPEs in each observation are more clearly presented in Figure \ref{fig:gsn069rec}. 
It is interesting to see that the TOA of both the observed and simulated QPEs are with alternative 
short and long intervals, which irregularly vary by amount of $2-3$ks. 
Such quasi-periodic behaviour can be explained by combination effects of the Schwarzschild precession and the presence of a small eccentricity of the orbit of the star.
{
The time interval between QPEs  should be in orders 
of $P_{1/2}+\delta P (f_0)$ where $P_{1/2}$ is the half orbital period and $\delta P (f_0)\sim 4e\sin f_0/\dot \omega_\bullet=4e\sin f_0(\bh G)^{-1/2} a_\bullet^{3/2}$ (when $e_\bullet\ll1$), where $\dot \omega_\bullet=(G \bh/a_\bullet^3)^{1/2}$ is the orbital angular velocity, $f_0$ is the true anomaly of the orbitor when it hits on the disc. If there is no orbital precession ($f_0={\rm constant}$), the 
time interval should be either $P_{1/2}+\delta P(f_0)$ or $P_{1/2}+\delta P(f_0+\pi)=P_{1/2}-\delta P(f_0)$. Due to Schwarzschild and Lense-thirring orbital precession, $f_0$ precess in each orbit and the interval of QPEs will be modulated between $P_{1/2}+\delta P(\pi/2)$ and 
$P_{1/2}-\delta P(\pi/2)$. For GSN069, we find that $\delta P(\pi/2)\sim2.4$ks, which is 
consistent with those shown in the right panel of Figure~\ref{fig:gsn069rec}. 
In Schwrzschild black hole, the orbital precession become $2\pi$ after in total of 
$N_{\rm p, S}$ flares, which is 
\be
N_{\rm p, S}=\frac{4\pi}{\delta \omega_{\rm S}}=\frac{2a_\bullet}{3r_g}(1-e_\bullet^2)\simeq 259, 
\ee
where $\delta \omega_{\rm S}=6\pi (a_\bullet/r_g)^{-1}(1-e_\bullet^2)^{-1}$.
This is also consistent with the period of modulation $N_{\rm p, S}\sim 258$ 
shown in the right panel of Figure~\ref{fig:gsn069rec}. 
The difference of time per flare is then approximately $2\delta P(\pi/2)/N_{\rm p, S}\propto a_\bullet^{1/2}$\footnote{See similar ``sidereal period'' in the case of Schwarzchild, Lense-thirring and quadruple momentum orbital precession in~\citet{2016MNRAS.460.2445I, 2017ApJ...839....3I}.}, thus the larger the distance of the orbitor, the larger the relativistic effects on TOA of QPEs around a Schwarzchild MBH. 
%E= acos e
%cos E= e ; sin E = + - (1-e**2)**0.5     E simeq pi/2 - e
%n1=pi/2-e - e * (1-e**2)*0.5=pi/2-2e
%n2=pi/2-e + e*(1-e**2)*0.5  =pi/2 + pi
%dt= d n1 / (mg/a**3)**0.5 =(pi - 2e) / dot omega
%4e/dot omega
%/dot omega=2pi/P=(mbh G/a**3)**0.5
%d_s omega = 6\pi mbh G/[c^2 a(1-e^2)]
% 3\pi / c^2 (mbh G a )**0.5 /(1-e^2)
}

As a summary, we find that the observed TOA of the QPEs of GSN069 can be well explained by 
the star-disc collisions models and some of the parameters of the orbit of the star can be well constrained by using only the timing of the X-ray emission of QPEs.
{ The quasi-periodic behaviour can be explained by combination effects of a small orbital eccentricity of the orbitor and the Schwarzschild precession.}

\section{ Is the orbitor in GSN069 a stellar core from a previous partial tidal disruption event?}
\label{sec:possible_pTDE}

{ 
Recent observations suggest that GSN069 exhibit some characteristic 
of partial tidal disruption event (TDE)~\citep{2021arXiv210903471Z,2021arXiv210901683S}.
Considering that the QPEs of GSN069 exhibit long-lived signatures of tidal disruption event~\citep{2018ApJ...857L..16S}, here 
we show that, if the star-disc collision model is the 
driven mechanism of QPEs of GSN069, the orbiter may be a stellar core remained after a partial tidal disruption of an evolved star (possibly a red giant) around the MBH in GSN069. More details are provided as follows.}

{
Partial TDEs can happen if a 
red giant star pass through the MBH near the tidal radius $r_p\sim r_t=r_\star (\bh/m_\star)^{1/3}$, 
where $r_\star$ and $m_\star$ is the radius and the mass of the red 
giant~\citep[e.g.,][]{2014ApJ...788...99B,2019ApJ...883L..17C,2021ApJ...914...69C,2012ApJ...757..134M}. 
Typically for red giant $r_\star=5\sim20R_\odot$ when $m_\star\sim1-5\msun$~\citep{2012ApJ...757..134M}.
As the stellar core is more compact than the outer envelope, the stellar core remained while the 
envelope is disrupted by the tidal force of the MBH. 
Then the stellar core should be initially with orbital periapsis $r_p\sim300-2075r_g$ 
for GSN069 with $\bh\sim3\times10^5\msun$. This result is consistent with the distance of 
$\sim400r_g$ of the orbitor from the MBH for GSN069 resulting from our constraints. 
Note that the above estimation is for initial value of $r_p$ and the orbit of the stellar core may evolved later 
due to the collisions with the accretion disc (See text later).}

%\item { Duration of the QPEs of GSN069. The duration of each flare is expected to be 
%the crossing time $t_D$ of the star in the accretion disc, which is expected to be given by 
%$t_{\rm D}=\frac{2h}{v_\bullet}$ where $h$ is the scale height of the disc and $v_\bullet$ 
%is the orbital velocity of the star. For GSN069 we expect $h\sim 10^6$km and 
%$v_{\bullet}=1.5\times10^4\kms$ if the orbitor is at $\sim400r_g$, thus the 
%expected time of crossing may be $t_{\rm D}\sim100$s. 
%$\sim$
%}
{ During each collision, part of the kinematic energy of the star are transferred into the 
shocked gas. As the radius of the remnant core is usually $r_{\rm core}\sim r_\star/20\sim0.1R_\odot$
\citep{2012ApJ...757..134M}, the X-ray luminosity of the eruption is then given by
\be\ba
L_X&\sim10^{43}~{\rm erg/s}~f_X\frac{\epsilon^2}{\lambda^2}\left(\frac{r_{\rm core}}{0.1~R_\odot}\right)^{2}
\frac{0.01}{\alpha}\left(\frac{\bh}{10^6~\msun}\right)^{-1},
\label{eq:lx}
\ea\ee
where $f_X$ is the fraction of emission in X-ray that depends on the details of radiative processes~\citep{2004A&A...413..173N}, $r_{\rm core}$, $\alpha$, $\epsilon$ and $\lambda$ is the radius of the stellar core, 
viscosity, radiative efficiency and Eddington ratio of the disc, respectively. 
The typical X-ray luminosity of QPEs are in orders 
of $10^{41}\sim10^{42}$ erg$/$s~\citep{2019GSN069QPE, 2021Natur.592..704A}, thus the above estimation
is consistent with the observed luminosity when $f_X\sim0.01-0.1$. 
}

{ The star will experience drag force caused by 
sweeping up of the intercepted disc gas in each collision, and finally the orbit is circularized 
and aligned with the disc. Following~\citet{1995MNRAS.275..628R}, the alignment timescale ($t_{\rm align}$) is assumed approximately the time needed for star to intercept with the disc material with masses equal to its own masses, which 
is given by \footnote{ Note that orbital eccentricity and inclination 
respect to the disc plane can also affect the alignment timescale 
(For more details see~\citet{1993MNRAS.265..365V, 1995MNRAS.275..628R}). If initially the orbital inclination is low, 
the  eccentricity will stay at low eccentric value for a long time~\citep{1995MNRAS.275..628R}. 
Thus, the probability of observing GSN069-like system with low eccentric value of the orbitor ($\sim0.06$) 
will be higher if initially the orbital inclination is low.}
\be\ba
\frac{t_{\rm align}}{t_{\rm orb}} &\sim 1.3\times10^6
\frac{m_{\rm core}}{0.1 \msun}\left(\frac{r_{\rm core}}{0.1~R_\odot}\right)^{-2}\left(\frac{\alpha}{0.01}\right)\\
&\times\left(\frac{\lambda}{\epsilon}\right) \left(\frac{a_{\bullet}}{10^2~r_{\rm g}}\right)^{-3/2},
\label{eq:t_align}
\ea\ee
where $t_{\rm orb}$ is the orbital period, $m_{\rm core}\sim 0.3-0.6\msun$~\citep{2012ApJ...757..134M} 
 is the mass of the stellar core to the MBH, respectively.  
For GSN069 with $\bh\sim3\times10^5\msun$ and $a_{\bullet}\sim 400r_g$ (as shown in Section~\ref{sec:res_GSN069}), 
it is then suggest that the QPEs can last for about $\sim1170$ yrs. The corresponding gravitational wave orbital decay is 
with timescale of $\ge2\times 10^6$yr if $e\le0.6$ and $m_{\rm core}\sim 0.3\msun$.
As the observation time of QPEs are typically in about $\la1$yr, 
it is justify that in our simulations we ignore the changes of the orbit due to both 
the star-disc collision and the gravitational wave orbital decay.}

{ As QPE sources are assumed results of partial TDE events, the event rate of
QPE sources ($R_{\rm QPE}$) is expected to be the same as those of partial TDE ($R_{\rm pTDE}$), the later 
of which are expected to be about $10\%$ of those normal TDE (disrupting a main-sequence star), i.e., 
$R_{\rm pTDE}\sim0.1R_{\rm TDE}\sim10^{-5}$ yr$^{-1}$ per galaxy (with $\bh<10^{7}\msun$)~\citep{2012ApJ...757..134M}. 
As the duration of QPE events can be $T_{\rm QPE}\sim1000$yr, the duty cycle of QPE sources is then expected to be 
around $R_{\rm QPE}T_{\rm QPE}\sim10^{-2}$, suggesting that 
QPEs may appear in every one of a hundred late type galaxies (with $\bh<10^{7}\msun$) at any given moment. 
It will be more easy to 
understand the frequency of QPE sources if we obtain the number ratio between the QPE sources and those of normal TDE:
\be
\mathscr{R}=\frac{R_{\rm QPE}}{R_{\rm TDE}}\times\frac{T_{\rm QPE}}{T_{\rm TDE}}\sim 10^2-10^3
\ee
where $T_{\rm TDE}\sim0.1-1$yr is the duration of normal TDE events. 
The above estimation suggest that theoretically the QPEs sources should be much more abundant than 
those of normal TDE sources. Currently there are only a few QPE sources revealed, possibly because the 
observation of them is quite expensive, and that not all of them  produce bright enough flares. 
}

\section{Constraining the spin parameters of the MBH by TOA of QPEs}
\label{sec:spin}
\begin{figure*}
\center
    \includegraphics[scale=0.4]{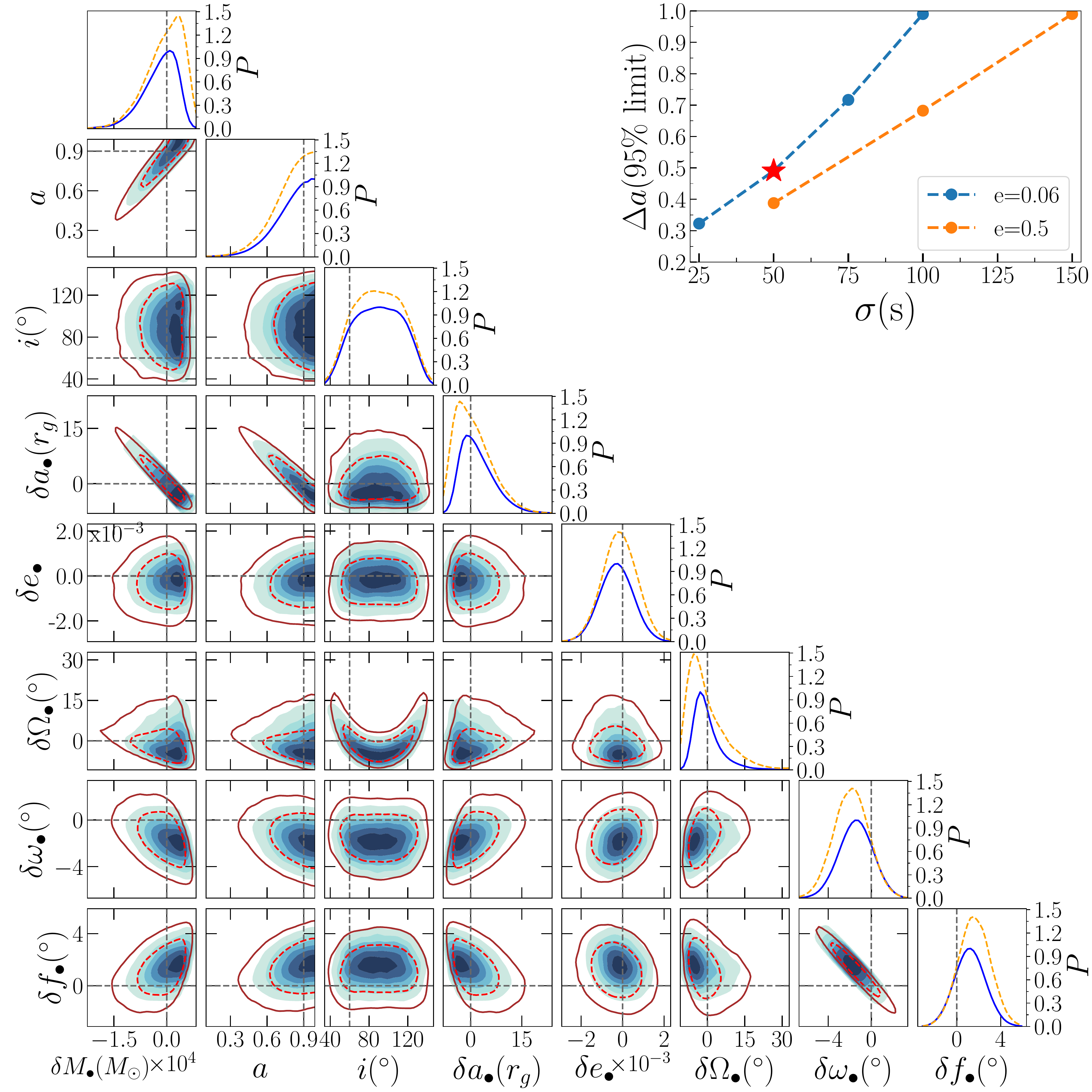}
\caption{Top right panel: Constraints on the amplitude of spin $a$ in the case of different 
orbital eccentricity, or the assumed timing measurement errors of 
the TOA of QPEs for a GSN069-like MBH. $25$ of $425$ QPEs are used for MCMC, 
corresponds to a duration of about $\sim 155$days.
Panels on the left: The constraints on the parameters in the case $\sigma_t=50$s and $e_\bullet=0.06$ 
(corresponds to the red star symbol in the top right panel). 
The contours and the marginalized probabilities are similar to those in 
Figure~\ref{fig:gsn069mcmccontour}. $\delta\bh$, 
$\delta a_\bullet$, $\delta e_\bullet$, $\delta \Omega_\bullet$, 
$\delta \omega_\bullet$, and $\delta f_\bullet$ show the difference respect to the input value of the 
mock data. The dashed line in each panel show the position of the input value.\label{fig:mock}}
\end{figure*}

%\begin{figure}
%\centering
%  \includegraphics[scale=0.26]{./spin_cons.pdf}
%  \caption{Constraints on the amplitude of spin $a$ in case of different orbital eccentricity, or 
%  the assumed timing measurement errors of 
%the TOA of QPEs for a GSN069-like MBH. $25$ of $425$ QPEs are used for MCMC, 
%corresponds to a duration of about $\sim 155$days. }
%   \label{fig:spincons}
%\end{figure}

The TOA of QPEs can be also affected by the presence of the spinning of the MBH due to 
Lense-Thirring effects on the orbit and the additional deflection of lights by the frame-dragging effects.
%We find that the spin parameter of GSN069 can not be reconstructed by the current observed timing of 
%QPEs in GSN069. However, we show here that the TOAs of QPEs can set tight constraints on 
%the spin parameters of the MBH, if the uncertainties on the timing of TOAs can be accurately enough for an observations that covers 
%the QPEs of a source across time scales of about several month. 
We investigate the constraints on the spin parameters of the MBH (also other model parameters simultaneously) 
by using mock observations, corresponds to which can possibly be collected in near-future for QPE sources. 
We find that the inclination $I_\bullet$ has a near-degeneracy with the spin parameter $a$, 
which largely slows down the convergence of the MCMC simulations. As usually $I_\bullet$ can 
only be poorly constrained, here we fix $I_\bullet$ for simplicity. If $I_\bullet$ is taken as 
a free parameter, we find that our following constraints on $a$ is only slightly weaker (e.g., 
increase from $\Delta a\sim0.5$ to $\Delta a\sim 0.6$).  

For demonstration purpose, we choose QPEs from a GSN069-like galaxy given the best-fit values 
suggested by Table~\ref{tab:mcmcpars}: $\bh = 2.65 \times 10^5 M_{\odot}$ (or $P_k=63.16$ks),
$a=0.9$, $i=60\degree$, $a_\bullet=390r_g$, $e_\bullet=0.06$ or $0.5$,
$I_\bullet=45^\circ$, $\Omega_\bullet=120\degree$, $\omega_\bullet=140\degree$ and 
$f_\bullet=160\degree$. 
There are in total of $8$ free parameters in the MCMC: $(P_k, a, i,a_\bullet, e_\bullet, 
\Omega_\bullet, \omega_\bullet, f_\bullet)$. 
We simulate in total of $425$ flares, which corresponds to a time duration of
$\sim155$ days. To mimic the intermittent observation, only the $1-5$th, $106-111$th, $\cdots$ 
and $421-425$th eruptions (in total of $25$ eruptions) are used for recovering. 

By performing a number of MCMC simulations, we explore the constraints on the spin 
magnitude in case of different timing measurement uncertainties or the orbital eccentricities. 
The results are shown in Figure~\ref{fig:mock}. In the case when $e_\bullet=0.06$, the constraints of spin are only possible if {$\sigma_t\la 100$}s. 
However, if $e_\bullet=0.5$, the constraints can be stronger and the 
spin parameter can be revealed if {{$\sigma_t\le150$}}s. 
The results of the reconstructed model parameters are also shown in the left panels in Figure~\ref{fig:mock}
given the uncertainties of TOA $\sigma_t = 50s$ and $e_\bullet=0.06$. 
We can see that both the orbital elements and the spin of MBH can be recovered, 
where the uncertainties of {$\Delta a\sim 0.49$} and {$\Delta i\sim 85\degree$}, respectively.

In principle, orbiters similar to the above one but with higher eccentricities, 
or QPE observations covering longer time span, can set even stronger constraints on the 
spinning of MBH. { By methods similar in Section~\ref{sec:res_GSN069}, we expect that the 
time difference per flare due to the lense-thirring effect is approximately 
$2\delta P(\pi/2)/N_{\rm p, L}$, 
where $N_{\rm p,L}=4\pi/|\delta\omega_L|$, and $|\delta \omega_L|=12\pi a (a_\bullet/r_g)^{3/2}(1-e_\bullet^2)^{-3/2}\cos I_\bullet$. Note that $2\delta P(\pi/2)/N_{\rm p, L}$ is independent of 
$a_\bullet$. Similarly, the variations due to quadruple momentum effects $2\delta P(\pi/2)/N_{\rm p, Q}\propto a_\bullet^{-1/2}$, where $N_{\rm p, Q}\propto a_\bullet^{-5/2}$}. We defer exploring more details of the constraining of the spin on various 
conditions of the orbitors by timing of QPEs to future studies.

\section{Summaries}
X-ray quasi-periodic eruptions (QPEs) in active galactic nuclei are found recently with 
unknown driven mechanisms. 
By assuming that QPEs are driven by star-disc collisions, 
we develop a full relativistic model to reconstruct the parameters of the orbiter and also
the spinning of massive black hole (MBH) by using the time of arrival (TOA) of these QPEs. 

We apply our method to the QPEs of GSN069, and find that the orbital parameters of the star can be well constrained, although the spinning of MBH can not, due to the limited  timing accuracy of 
the observed samples. The orbiter in GSN069 is found with semimajor axis of about {$365^{+54}_{-49}r_{\rm g}$}
and small eccentricity ({$e_\bullet=0.05^{+0.02}_{-0.02}$}) around a MBH with {$\bh=3.0^{+0.9}_{-0.6}
\times10^5\msun$}, 
We find that the variations on the recurring time of QPEs of GSN069, 
with repeating irregular alternative long and short intervals, 
can be well explained by the presence of the orbital eccentricity and the relativistic orbital precession.
{ We also discuss the possibilities that the QPEs in GSN069 are caused by continual 
collisions between a stripped stellar core with the accretion disc around the MBH, the former of which are formed by partial tidal disruption of a red giant.}

We find that, for a GSN069-like QPE source
that last more than several month, if the timing uncertainties of the TOA of QPEs can be less 
than {$\sim 100-150$} seconds, the spinning of the MBH can be constrained. 
Orbitors with higher eccentricities, or QPE observations covering longer time span, can be used to 
set even stronger constraints on the spinning of MBH.

If the X-ray QPEs are driven by the star-disc collisions around the MBH, 
the time of arrival of each QPEs includes various general relativistic effects
around the MBH. Our results suggest that long-time monitoring of X-ray QPEs can provide 
a unique probe of  general relativistic effects, the spinning of MBH, tests of the 
no hair theorem, and also gravity theories.

\acknowledgments
 We thank the anonymous referee for the helpful comments that have improved this paper.
This work was supported in part by the Natural Science Foundation of Guangdong
Province under grant No. 2021A1515012373, National Natural Science Foundation of 
China under grant No. 11603083, U1731104. This work was also supported in part by
the Key Project of the National Natural Science Foundation of China under grant No. 11733010, 12133004.  
The simulations in this work are performed partly in the TianHe II National
Supercomputer Center in Guangzhou.

% \bibliography{sdcol}{}

\end{document}